\documentclass[titlepage]{article}
\usepackage{fullpage}
\usepackage{amsmath}
\usepackage{amssymb}
\usepackage{orcidlink}
\usepackage{hyperref}

\begin{document}

\title{Total Loss Functions for Measuring the Accuracy of Nonnegative Cross-Sectional Predictions}
\author{Charles D. Coleman
\orcidlink{0000-0001-6940-8117}\thanks{ This paper reports the general results of research originally undertaken while the author was employed by the Census Bureau.  The views expressed are attributable to the author and do not necessarily reflect those of the Census Bureau.}\\
Timely Analytics, LLC \\
E-mail: info@timely-analytics.com}

\maketitle

\abstract{The total loss function associated with a set of cross-sectional predictions, that is, estimates or forecasts, summarizes the set’s overall accuracy. Its arguments are the individual cross-sectional units’ loss functions.  Under general assumptions, including impartiality, about the forms of the individual loss functions, and the specific assumptions that the total loss function is anonymous and monotonic, only the additive, multiplicative and L-type (with restrictions) total loss functions are found to be admissible. The first two total loss functions correspond to different interpretations of economic utility.  An isomorphism exists between these two total loss functions. Thus, the additive total loss function can always be used. This isomorphism can also be used to explore the properties of various combinations of total and individual loss functions.  Moreover, the additive loss function obeys the von Neumann-Morgenstern expected utility axioms.}

\newpage

\section{Introduction}

Various measures have been proposed to assess the accuracy of sets of cross-sectional predictions.  These measures can often be thought of as total loss functions which aggregate into a single measure individual loss functions, whose arguments are the actual and estimated values for each cross-sectional unit.\footnote{Spencer (1986) refers to the individual and total loss functions as the component loss and loss functions, respectively.} Commonly used examples includemean absolute percentage error, median absolute percentage error, mean squared error and root mean squared percentage error.\footnote{Coleman (2025), Section 5, has a typology of these measures.} Under mild assumptions about the individual loss functions, including impartiality, and the assumptions of anonymity and monotonicity of total loss, only the additive, multiplicative and $L$-type total loss functions,\footnote{The last is with qualifications.  See Corollary 1 and Theorem 1 in Section 3 below.} and their positive monotonic transformations are shown to be admissible. Furthermore, a fundamental isomorphism exists between these two first two classes of total loss functions, so that only one class need be used. Since the additive total loss function has a simple interpretation using expected utility theory, it is preferred. This isomorphism can then be used to explore properties of accuracy measures. Finally, all of these total loss functions are shown to obey Fisher-consistency: they are minimized when the estimates equal the true values.

Section 2 introduces and makes assumptions about individual loss functions. Section 3 does the same for total loss functions. Section 4 offers interpretations of the additive and multiplicative total loss functions. Section 5 proves the isomorphism between these total loss functions and uses this isomorphism to examine the properties of a particular accuracy measure.  Section 6 proves Fisher-consistency.  Section 7 concludes this paper.

\section{Individual Loss Functions}

Let there be  $n$ individual cross-sectional units with predictions $P_i$ and actual values $A_i$ . Each pair $(P_i ,A_i )$ has an
associated individual loss function $L_i(P_i,A_i)$.  The total loss function is a function of all of the individual loss functions:
$\mathcal{L}= f[L_1(P_1,A)1),\ldots,L_n(P_n,A_n)]$.

We first make assumptions on the $L_i$:\newline
\textbf{Assumption 1 (unrestricted domain):} For all $i$, and real $P_i$, and $A_i$ in the domain of interest,  $L_i(P_i,A_i)$ exists and is
single-valued.  In particular, $L_i(A_i,A_i)$ is defined.\newline
Assumption 1 makes $L_i$ well-defined for every real pair $(P_i,A_i)$ in (at least) a subset on the real plane  $\Re^2$  and is thus a
function on its domain. The domain of interest depends on the problem at hand. In the case of income predictions, the entire real plane may be the domain.  Population predictions are always nonnegative,\footnote{They may be 0 in unpopulated areas, but these are usually discarded from analysis.} so the domain is the nonnegative
quadrant of  $\Re^2$, that is, $\Re^2_+$. This domain is bounded below by 0 on both axes.  Similar upper bounds may exist.

\textbf{Assumption 2 (boundedness):} $0 < L_i(P_i,A_i) < \infty$ for all $i$ and all finite $P_i, A_i$.\newline
Assumption 2 prevents It forces the individual losses to be nonnegative, so as to preserve order in the multiplicative total loss function. It also comes into play in the fundamental isomorphism, Theorem 2, below.  It also prevents the total loss function $\mathcal{L}$, defined below, from becoming unbounded when all of its arguments are finite. 

\textbf{Assumption 3 (monotonicity):} For all $P'_i$ such that either $P'_i < P_i  \le A_i$ or  $P'_i > P_i  \ge A_i$ ,
$L(P'_i,A_i) >L(L(P_i,A_i)$.\newline
Assumption 3 assures that the individual losses increase in the distance of the estimate from the actual value. Together, Assumptions 1-3 make the individual loss functions well-behaved.  

Finally, we assume the loss functions are identical:\newline
\textbf{Assumption 4 (impartiality):}
For all $P, A$ and all $j \ne i$, $L_i(P,A) = L_j(P,A) = L(P,A)$.\newline
The importance of Assumption 4 is that the identity of any unit $i$ is irrelevant to calculating its loss. All units are treated equally.

\section{The Total Loss Function}

Having characterized the individual loss functions up to impartiality, the final assumptions needed for the total loss function are anonymity and monotonicity of total loss:\newline
\textbf{Assumption 5 (anonymity):} 
Let $\mathcal{L} = f[L(P_1,A_1),...,L(P_n,A_i)]$ and let  $\mathcal{L'} = f[L'(P_1,A_1),...,L'(P_n,A_i)]$\newline
where the
arguments of $\mathcal{L'}$ are any permutation of the arguments of $\mathcal{L}$. Then,  $\mathcal{L'} = \mathcal{L}$.\newline
Assumption 5 is similar to Assumption 4 in that it guarantees equal treatment of all units. This time, the identity of any unit $i$ has no effect on the computation of total loss. 

Assumption 6 makes $\mathcal{L}$ “responsive” to every change in the individual loss functions.\newline
\textbf{Assumption 6 (monotonicity of total loss):} Assume  $\mathcal{L}$,  $\mathcal{L'}$ per Assumption 5.  Let $L(P'_i,A_i) > L(P_i,A_i)$ for one
$i$ and  $L(P'_j,A_j) > L(P_j,A_j)$ for all $j \ne i$. Then, $\mathcal{L'} > \mathcal{L}$.

The definitions of the various total loss functions are now in order.\newline
\textbf{Definition 1:} The arithmetic total loss function is $\mathcal{L}_A = \sum_{i=1}^n L(P_i,A_i)$.\newline
\textbf{Definition 2:} The multiplicative total loss function is $\mathcal{L}_M = \prod_{i=1}^n L(P_i,A_i)$.\newline
\textbf{Definition 3:} The quantile total loss function is $\mathcal{L}_Q = L_q \mid (\textrm{the proportion of }L_i \le L_q) = q$. 
When $q = 1$,
this becomes the maximum total loss function $\mathcal{L}_{MAX} = \max_i L_i$.\newline
\textbf{Definition 4:} The $L$-type total loss function is $\mathcal{L}_L = \sum_{i=1}^n c_i L_{(i)}(P_i,A_i)$ where the $c_i \ge 0$ are fixed coefficients and the $L_{(i)}$ are the losses sorted in either ascending or descending order.\newline
\textbf{Definition 5:} The $L_+$-type total loss function is an $L$-type total loss function where the $c_i > 0$. 
This nonnegativity condition is
similar to Assumption 2, to assure the nonnegativity of $\mathcal{L}_{L_+}$.

It can easily be seen that every quantile total loss function is an $L$-type total loss function by setting $c_i = 1$ for
the appropriate quantile and $c_j = 0$ for all $j \ne i$. Likewise, the additive total loss function is an $L$-type total loss function
with all $c_i = 1$ (or any other constant, see Theorem 1 below).

With these definitions in hand, it is time to characterize the total loss functions. Lemma 1 shows that
 $\mathcal{L}_Q$ violates 
Assumption 6. From here on, we will use the notation $L_i = L(P_i,A_i)$.\newline
\textbf{Lemma 1:} $\mathcal{L}_Q$ violates Assumption 6.\newline
\textbf{Proof:} Assume that the $L_i$ are sorted per Definition 4 , supressing the parentheses.  Case 1: $q < 1$. Let $L_{max} = \max_i L_i$. Clearly, $L_{max} > L_q = \mathcal{L}_Q$.  Now, let  $L'_{max} = L_{max} + \delta, \delta > 0$. $L_q$ is 
unchanged and, therefore, $\mathcal{L}_Q$ is unchanged.  Thus, Assumption 6 is violated.\newline
Case 2: q = 1. In this case, 
$\mathcal{L}_Q =  \mathcal{L}_{MAX} = L_{max}$. Now, let  $L'_i =L_i + \epsilon$, where $L_i < L_{max}$ and  $\epsilon > 0$ so that  $L'_i < L_{max}$ . By continuity of the
real numbers, this is always possible.  Similar to Case 1, $L_{max}$ and  $\mathcal{L}_{MAX}$ are unchanged and Assumption 6 is violated.$\square$

Lemma 1 has an important corollary.\newline
\textbf{Corollary 1:}  $\mathcal{L}_L$ with at least one $c_i = 0$ violates Assumption 6.\newline
\textbf{Proof:} Similar to Lemma 1.$\square$

The importance of Lemma 1 is that quantile loss functions with $q < 1$ are completely insensitive to the most
extreme outliers.\footnote{This formalizes Coleman’s (2025) criticism of quantile loss functions on the basis of extreme insensitivity to outliers.  Coleman (2025) also points out that these violate von Neumann-Morgenstern expected utility.}	$\mathcal{L}_{MAX}$ on the other hand, has some importance, given its relationships to game theory and statistical
decision theory. Minimizing  
$\mathcal{L}_{MAX}$ is equivalent to finding the minimax set of predictions. Corollary 1 effectively rules out
all $L$-estimators in common use, such as trimmed and Winsorized means.

Assumptions 5 and 6 lead to the following theorem, in which any transformations applied to $L$ are done before computing total loss:\newline
\textbf{Theorem 1:} The only total loss functions that satisfy Assumption 5, up to positive monotonic transformations,
are the additive, multiplicative and $L$-type with all $c_i > 0$ total loss functions:
 $\mathcal{L}_A$, $\mathcal{L}_M$, and $\mathcal{L}_{L_+}$, the $L_+$-type total loss function.\newline
 \textbf{Proof:} Let $I$ be the identity operator.  Assumption 5 implies that $I(L_1 \circ L_2) = I(L_2 \circ  L_1)$, where $\circ$ represents an operator. That is, the operator is commutative. The only arithmetic operators which satisfy this property are addition ($+$) and multiplication
($\times$).  In addition, the permutation operator applied to two elements is also commutative.\newline
(i) Substituting the addition and multiplication operators in the preceding shows that $\mathcal{L}_A$ and $\mathcal{L}_M$ satisfy Assumption 5.\newline
(ii) Noting that $c_i L_i$ is constant for all $i$, given $A_i, P_i$ reduces the case for $\mathcal{L}_{L_+}$  to that of $\mathcal{L}_A$, already proven.$\square$\newline
\textbf{Theorem 2:} The only total loss functions that satisfy Assumption 6, up to positive monotonic transformations,
are the same as those in Theorem 1.\newline
 \textbf{Proof:} Let $L'_i > L_i$ for some $i$.  In each case, $\mathcal{L'} > \mathcal{L}.\square $

Theorem 3 combines the results of Theorems 1 and 2:\newline
\textbf{Theroom 3:} The only total loss functions that satisfy Assumptions 5 and 6, up to positive monotonic transformations,
are the additive, multiplicative and $L$-type with all $c_i > 0$ total loss functions:
$\mathcal{L}_A$, $\mathcal{L}_M$, and $\mathcal{L}_L$ with all $c_i > 0$.

Theorem 3 itself has several corollaries.\newline
\textbf{Corollary 2:} Any monotonically increasing function $g$ of $\mathcal{L}_A$, $\mathcal{L}_M$ or $\mathcal{L}_{L_+}$,
 generates an order isomorphism.\newline
\textbf{Proof:}  Let $h$ be an monotonically increasing function. Let  $\mathcal{L'} > \mathcal{L}$. Then, $g(\mathcal{L'} )> g(\mathcal{L}$) and these values are unique for each value of  $\mathcal{L}$. The converse is also true. Therefore, $g$ is an order isomorphism.$\square$

The importance of Corollary 2 is that operations such as taking means, exponents and logarithms of the total loss function preserve its rankings. The transformations can also be iterated.

\section{Interpretating the Total Loss Function}

The two total loss functions $\mathcal{L}_A$ and $\mathcal{L}_M$
can be interpreted using utility theory. We begin by inverting loss to obtain utility,
using the appropriate inverse, additive or multiplicative. In the case of the additive total loss function $\mathcal{L}_A$, $U_i=-L_i$ where $U_i$
is unit $i$’s utility. We can thus reexpress $\mathcal{L}_A$
 as $\mathcal{L}_A = \sum_{i=1}^n L_i = \sum_{i=1}^n -U_i = -\sum_{i=1}^n U_i$.\footnote{In the general case, $U_i = a - bL_i, b > 0$. However, including the constants complicates the argument unnecessarily.} Therefore, total loss is the additive
inverse of the sum of the individual utilities. This corresponds to the case of an individual decision-maker’s evaluating a set of estimates.\footnote{See Coleman (2025) for a further discussion of this individual decision-maker.} The decision-maker has von Neumann-Morgenstern (1944) expected utility. That is, given a gamble
with two possible values of the utility of an estimate, $U^1$ and $U^2$, with respective probabilities $p$ and $1 - p$, the expected
utility of the gamble is $pU^1 + (1 - p)U^2$. In ranking sets of estimates, the set with the lowest sum of losses has the greatest
utility to the decision-maker.

In  the  case  of  the  multiplicative  total loss function $\mathcal{L}_M$,
we take  $U_i = L_i^{-1}$. We reexpress $\mathcal{L}_M$ as $\mathcal{L}_M = \prod_{i=1}^n L_i = \prod_{i=1}^n U_i^{-1} = 1/\prod_{i=1}^n U_i$. Thus,
$\mathcal{L}_M$ is the reciprocal of the product of the utilities. Minimizing total loss
is equivalent to maximizing the product of the individual utilities. The maximum product of the utilities is known as the Nash Bargaining Solution (Nash, 1950) with status quo utility equal to zero.\footnote{In the general case, the Nash Bargaining Solution maximizes the product  $\prod_{i=1}^n (U_i - U_i^*)$, where $U_i^*$  is agent $i$’s status quo utility. Taking $U_i^* = 0$  for all $i$ can be justified by noting that utility decays to zero as error and thus, loss, diverges.} This solution maximizes utility for each unit, subject to constraints. Each unit’s utility is assumed to be von Neumann-Morgenstern. Rubinstein et al. (1992) have generalized the Nash Bargaining Solution to non-von Neumann-Morgenstern frameworks. Their interpretation is based on bargaining and is beyond the scope of the present paper.

The importance of Assumptions 4 and 5 is now revealed. Each unit is treated identically in the computation of total loss.  Impartiality (Assumption 4) means that each individual’s preferences are independent of his identity and
actual value. This he great difference between
 $\mathcal{L}_M$ and the Nash Bargaining Solution is that, in the latter, individuals
are allowed to possess different preferences, while, in the former, all units are assumed to have the same preferences. Anonymity (Assumption 5) implies that the identity of the individual does not matter in computing total loss. Changing any individual’s loss function alters the rankings of total losses generated by different sets of estimates.\footnote{Note that multiplying any individual’s loss function by a positive constant has no effect on the rankings
generated by $\mathcal{L}_M$.  However, the rankings of the isomorphic $\mathcal{L}_A$ may be changed.  This is a consequence of the
violation of Assumption 4.}

\section{The Fundamental Isomorphism \label{FI}}

This Section proves that the additive and multiplicative total loss functions 
$\mathcal{L}_A$ and $\mathcal{L}_M$ are order isomorphic. The upshot is that
either total loss function can always be used.  Normally,
$\mathcal{L}_A$ is used, such as in the Mean Absolute Percentage Error
(MAPE), Root Mean Squared Error (RMSE) and Coleman’s (1999) loss function framework. Theorem 4 proves the isomorphism.\newline
\textbf{Theorem 2:}  $\mathcal{L}_A$ and $\mathcal{L}_M$ are isomorphic isomorphic by \newline
\textbf{Proof:}  See Beachy and Blair (2006, p. 129) and note that it is true for any base $b > 1$. Moreover, order is preserved by the exponential and logarithm functions, making this an order isomorphism.$\square$\newline
In practice, this isomorphism, accomplished in part by taking logarithms, can produce values of $\mathcal{L}_A < 0$ with at least some $L_i \le 0$.  This can be remedied by adding a constant $k$ to those values per Corollary 2 to make each  $L_i > 0$.  This is equivalent to multiplying each $L_i$ in $\mathcal{L}_M$ by a constant that makes each $L_i > 1$, also permitted by Corollary 2.

One power of Theorem 2 lies in identifying degeneracies in isomorphic loss functions.  For example, consider
the geometric mean absolute percentage error (GMAPE) (Swanson, Tayman and Barr, 2000): $\textrm{GMAPE} = \left(\prod_{i=1}^n \textrm{APE} _i \right)^{1/n} =  \left(\prod_{i=1}^n 100 |P_i - A_i| / A_i \right)^{1/n} = 100 \left(\prod_{i=1}^n |P_i - A_i| / A_i \right)^{1/n}$. By Theorem 2, we
can find the isomorphic additive total loss function $\mathcal{L}_A = \sum_{i=1}^n \log \textrm{APE} _i$ (again the base is irrelevant). If for
any $i$, $P_i = Ai_i$, that is, the estimate for unit $i$ is perfect,  $\textrm{APE} _i = 0$, its logarithm becomes $-\infty$ both
violating Assumption 2 and making $\mathcal{L}_A = -\infty$. In this case,
GMAPE = 0, no matter what its other arguments are. This degeneracy prevents GMAPE from being a useful accuracy measure.

\section{Fisher-Consistency}

Spencer (1980, p. 36) introduced the concept that the total loss function should be Fisher-consistent: the total loss
function must be minimized when $P_i = A_i$ for all $i$. To prove this is true of out total loss functions, we first need Lemma 2, which proves that $L(P_i$ 
is minimized with respect to $P_i$ when $P_i = A_i$.\newline
{\textbf Lemma 2:} $L(P_i,A_i)$ is minimized with respect to $P_i$i when $P_i = A_i$.\newline
\textbf{Proof:} Assume the contrary, then there exists $P'_i \ne A_i$ such that $L(P'_i,A_i) < L(P_i,A_i)$.  But this contradicts
Assumption 3.  Therefore, the Lemma is proved.$\square$

Theorem 3 then proves Fisher-consistency for  $\mathcal{L}_A, \mathcal{L}_M, \mathcal{L}_Q$ and $\mathcal{L} _{L_+}$.\footnote{However, note that Fisher-consistency does not imply usefulness.}\newline
{\textbf Theorem 3 (Fisher-consistency):} $\mathcal{L}_A, \mathcal{L}_M, \mathcal{L}_Q$ and $\mathcal{L} _{L_+}$ are minimized when $P_i = A_i$ for all $i$.\newline
\textbf{Proof:} We first observe from Lemma 2 that $L(P_i,A_i)$ is minimized with respect to $P_i$ when $P_i = A_i$. Since this is true for all $i$, the following claims are true:\newline
(i) $\mathcal{L}_A$:  $\mathcal{L}_A = \sum_{i=1}^n L(P_i,A_i)$ is minimized.\newline
(ii) $\mathcal{L}_M$: $\mathcal{L}_M = \prod_{i=1}^n L(P_i,A_i)$ is minimized, noting from Assumption 1 that $L(P_i,A_i) \ge 0$ and, therefore,  $\mathcal{L}_M \ge 0$.\newline
(iii) $\mathcal{L}_Q$:  $\mathcal{L}_Q = L_q$ is minimized.\newline
(iv) $\mathcal{L} _{L_+}$: The $L_{(i)}$ are minimized.  Therefore, $\mathcal{L}_{L_+}= \sum_{i=1}^n c_i L_{(i)}(P_i,A_i)$ is minimized.

\section{Conclusion}

This paper has developed total loss functions for measuring the cross-sectional predictions to satisfy several reasonable conditions. The most important conditions are impartiality of individual losses and anonymity and monotonicity of total loss. Under these conditions, only the additive and multiplicative and, theoretically, $L$-type total loss functions are admissible. Moreover, the first two are isomorphic, so that only one of the two can always be used. This isomorphism has been used to illustrate the source of the degeneracy of GMAPE. The additive total loss function further obeys the von Neumann-Morgenstern expected utility axioms, easing its interpretation relative to the multiplicative total loss function.  Finally, all of these total loss functions have been shown to be Fisher-consistent.

\section{References}

Beachy, John A.  and William D. Blair (2006), \textit{Abstract Algebra}, Third Edition, Waveland Press, Long Grove, Illinois.\newline\newline
Coleman, Charles, D., (2025) “Metrics for Assessing the Accuracy of Nonnegative Cross-Sectional Predictions,” \url{https://doi.org/10.48550/arXiv.2505.18130}\newline\newline
Gerber, Anke (1995) “The Nash Solution as von Neumann-Morgenstern Utility Function on Bargaining Games,” Working-Paper 244, Institute of Mathematical Economics.\newline\newline
von Neumann, John and Morgenstern, Oskar, (1944) \textit{Theory of Games and Behavior}, Princeton University Press, Princeton. Second edition, 1947; third edition 1953: Princeton University Press, Princeton. Reprinted 1963 and 1967: John Wiley and Sons, New York.\newline\newline
Jureèková, Jana and Kumar Sen, Pranab, (1996) \textit{Robust Statistical Procedures}, John Wiley and Sons, New York.\newline\newline
Nash, John F., (1950) “The Bargaining Problem,” \textit{Econometrica} \textbf{18}, 155-162.\newline\newline
Roth, John F., (1978) “The Nash Solution and the Utility of Bargaining,” \textit{Econometrica} \textbf{46}, 587-594.\newline\newline
Rubinstein, Ariel, Safra, Zvi and Thomson, William, (1992) “On the Interpretation of the Nash Bargaining Solution and Its Extension to Non-expected Utility Preferences,” \textit{Econometrica} \textbf{60}, 1171-86.\newline\newline
Spencer, Bruce (1980) \textit{Benefit-Cost Analysis of Data Used to Allocate Funds}, Springer-Verlag, New York.\newline\newline
Spencer, Bruce (1986) “Conceptual Issues in Measuring Improvement in Population Estimates,” in Bureau of the Census,
\textit{Second Annual Research Conference: Proceedings} March 23-26, 1986, 393-407.

\end{document}